		    \documentclass[11pt]{article}

\usepackage[body={16cm,23cm}]{geometry}

\usepackage{cite}
\usepackage{amsmath,amssymb}
\usepackage{amsfonts}
\usepackage{epic}
\usepackage{eepic}
\usepackage{enumerate}
\usepackage{graphicx}
\usepackage[hang,small,bf
]{caption}
\newcounter{app}
\newcounter{sapp}[app]

\newcommand{\ds}{\displaystyle}

\newcommand{\hf}{{\textstyle \frac{1}{2}}}

\newcommand{\s}{\sigma}

\def\bea{\begin{eqnarray}}
\def\eea{\end{eqnarray}}

\def\be{\begin{equation}}
\def\ee{\end{equation}}

\renewcommand{\author}[1]{\large\rm #1\\ \bigskip}

\renewcommand{\title}[1]{\bigskip\bigskip\Large\bf #1\bigskip\bigskip\\}

\begin{document}

\vglue 2cm

\begin{center}

\title{
Variational approach to the scaling function of the 2D Ising model
in a magnetic field
}

\vspace{1cm}

\author{Vladimir~V.~Mangazeev\footnote{%
Corresponding author: vladimir@maths.anu.edu.au}${}^{,a}$, 
Murray~T.~Batchelor${}^{\,a,b}$,
  Vladimir~V.~Bazhanov${}^{\,a}$ and Michael~Yu.~Dudalev${}^{\,a}$ }

{\it ${}^{a}$Department of Theoretical Physics,\\
         Research School of Physics and Engineering,\\
${}^{b}$Mathematical Sciences Institute,\\
    Australian National University, Canberra, ACT 0200, Australia.\\}

\end{center}

\vspace{1cm}

\begin{abstract}
The universal scaling function of the square lattice Ising model in a
magnetic field is obtained numerically via Baxter's variational corner
transfer matrix approach. The high precision numerical data is in perfect
agreement with the remarkable field theory results obtained by Fonseca and
Zamolodchikov, as well as with many previously known exact and numerical
results for the 2D Ising model. This includes excellent agreement with
analytic results for the magnetic susceptibility obtained by Orrick,
Nickel, Guttmann and Perk. In general the high precision of the numerical
results underlines the potential and full power of the variational corner
transfer matrix approach.
\end{abstract}

\newpage
The Ising model has played a prominent role in the
development of the theory of phase transition and critical phenomena
\cite{O44,O49,Y52,McCoyWu,Baxterbook,AF83,BPZ84,Zam89a}.
The partition function of the nearest-neighbour Ising model on the square 
lattice reads 
\be 
{Z}=\sum_{\sigma}\exp \Big\{\,\beta
\sum_{\langle ij\rangle} \sigma_i
  \sigma_j 
+{H}\sum_i \sigma_i\,\Big\}\ , \qquad \sigma_i=\pm1,\label{Z-def} 
\ee
where the first sum in the exponent is taken over all edges, the second over
all sites and the outer sum over all spin configurations
$\{\s\}$ of the lattice. The constants $H$ and $\beta$ denote
the (suitably normalized) magnetic field and inverse temperature. 
The specific free energy, magnetization and magnetic susceptibility 
are defined as 
\be
F=-\lim_{N\to\infty}\frac{1}{N}\log Z,\qquad
M=-\frac{\partial F}{\partial H},\qquad \chi=-\frac{\partial^2
  F}{\partial H^2}\ ,\label{F-def}
\ee 
where $N$ is the number of lattice sites. 
The model exhibits a second order phase transition at 
$\beta=\beta_c$, $H=0$, where \cite{O44}
\be
\beta_c=\frac{1}{2}\log(1+\sqrt{2})=0.44068679\ldots\ .\label{beta-c} 
\ee
In what follows we will exclude the temperature variable $\beta$ in favour 
of a new variable 
\be
2\tau={\rm cosech}(2\beta)-\sinh(2\beta)\ ,\qquad \tau_c=0,
\ee
which is vanishing for $\beta=\beta_c$ and positive
for $\beta<\beta_c$ (above the critical temperature).
Note also that this variable changes sign under the Kramers-Wannier
duality transformation for $H=0$. 
Another useful related variable is
\be
k=k(\tau)=(\sqrt{1+\tau^2}+\tau)^2.\label{k-def}
\ee
 
According to the scaling theory \cite{Pat79,Car96,AF83}, 
the leading singular part, $F_{sing}(\tau,H)$, of the free energy
\eqref{F-def} in the vicinity of the critical point
can be expressed through a universal function ${\mathcal F}(m,h)$,
\be
F_{sing}(\tau,H)={\mathcal F}(m(\tau,H),h(\tau,H)), 
\qquad \tau\to 0, \quad H\to0,\label{fsing}
\ee 
where $\tau$ and $H$ enter the RHS only through  
non-linear scaling variables \cite{AF80}, 
\be m=m(\tau,H)=-\sqrt{2}\,\tau +O(\tau^3)+O(H^2)+\ldots,\quad
h=h(\tau,H)=C_h \,H+H\,O(\tau)+O(H^3)\ldots,\qquad \label{mh-def}
\ee 
which are analytic functions of $\tau$ and $H$.
The coefficients in these expansions (for instance, the leading
coefficients $-\sqrt{2}$ and $C_h$) are specific to the square lattice
Ising model, however, the function ${\mathcal F}(m,h)$ 
is the same for all models in the 2D Ising model universality class. 
It can be written as 
\be
{\mathcal F}(m,h)=\frac{m^2}{8\pi} \, \log
m^2+
h^{16/15} \,\Phi(\eta), 
\qquad \eta=\frac{m}{h^{8/15}}\label{Phi-def}
\ee
where $\Phi(\eta)$ is a universal scaling function of 
a single variable $\eta$ (the scaling parameter).

The function ${\mathcal F}(m,h)$ has a concise interpretation in 
terms of 2D Euclidean quantum field theory. Namely, 
it coincides with the vacuum energy density of 
the ``Ising Field Theory'' (IFT) \cite{FZ03}. The latter is defined as a
model of perturbed conformal field theory with the action
\be
{\cal A}_{\rm IFT} = {\cal A}_{(c=1/2)} + \frac{m}{2\pi}\,\int\,
\epsilon(x)\,d^2 x +
h\,\int \,\sigma(x)\,d^2 x\,,\label{IFT}
\ee
where ${\cal A}_{(c=1/2)}$ stands for the action of the $c=1/2$ CFT 
of free massless Majorana fermions, $\sigma(x)$ and
$\epsilon(x)$ are primary fields of conformal dimensions $1/16$ 
and~$1/2$.
Their normalization is fixed by the usual CFT convention,
\be
|x|^2 \,\langle \epsilon(x)\epsilon(0)\rangle \to 1\,; \qquad 
|x|^{1/4}\,\langle \sigma(x)\sigma(0)\rangle \to 1\, \quad {\rm as}
\quad |x| \to 0\,.\label{norm}
\ee
With this normalization, the parameters $m$ and $h$ have 
the mass dimensions $1$ and $15/8$, respectively,
and the scaling parameter $\eta$ in \eqref{Phi-def} is dimensionless.

The scaling function \eqref{Phi-def} is of much interest as it controls 
all thermodynamic properties of the Ising model in the critical
domain. 
Although there are many exact 
results (obtained through exact solutions of \eqref{IFT} 
at $h=0$ and all $\tau$ 
\cite{O44, Kaufman, Y52, BMW, TM73, McCoy76}, and at $\tau=0$ 
and all $h$ \cite{Zam89a, Fateev94,DM95,WNS92,WPSN94,BNW,BS97}; these data
are collected in \cite{Delfino2004}) as 
well as much numerical data \cite{Ess68, Zin96, VicC,VicA,VicB}
about this function, its complete analytic characterization is still lacking. 

Recently \cite{FZ03} the function \eqref{Phi-def}, particularly its
analytic properties, have been thoroughly
studied in the framework of the Ising Field Theory \eqref{IFT}.  
The authors of \cite{FZ03} made extensive numerical
calculations of the scaling function $\Phi(\eta)$ 
using the ``Truncated Free-Fermion Space Approach'' (TFFSA), which is
a modification of the well-known ``Truncated Conformal Space
Approach'' (TCSA)\cite{YZ90,YZ91a}.

\begin{figure}[h]
\centering
\includegraphics[scale=1.0]{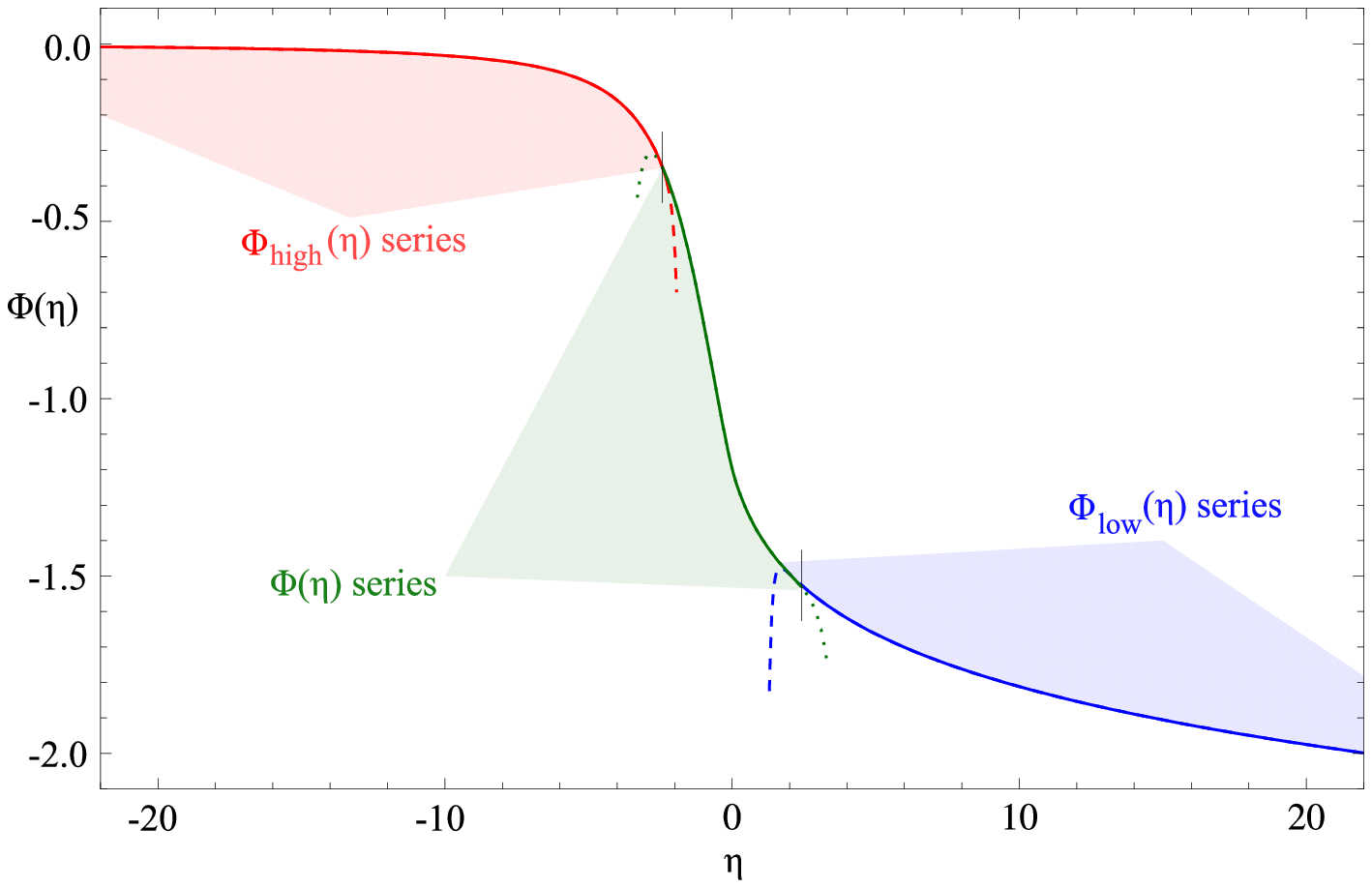} 
\caption{Scaling function of the two-dimensional Ising model in a
  magnetic field. The figure shows about 10,000 data points for the
  scaling function $\Phi(\eta)$. As expected, the points collapse on a
  smooth curve (their spread is much smaller than 
the resolution of the picture). The figure also shows plots of
the asymptotic expansions \eqref{Phi-low}, \eqref{Phi-high} and
\eqref{Phi-ser} with the coefficients given in Table~1, Table~2 and Table~3.} 
\label{figura1}
\end{figure}

The primary motivation of our work was to confirm and extend 
the field theory results of ref. \cite{FZ03} through {\em ab initio}
calculations, directly from the original lattice formulation 
\eqref{Z-def} of the Ising model. 
We used Baxter's variational
approach based on the corner transfer matrix method
\cite{BaxDim,BaxVar,Baxter2007}.  
The
main advantage of this approach over other numerical schemes (e.g., the
row-to-row transfer matrix method) is that it is formulated 
directly in the limit of an {\em infinite} lattice. 
Its accuracy depends on the magnitude of truncated eigenvalues of the corner
transfer matrix (which is at our control), rather than the size of the
lattice.  
The details of our calculations along with numerical data for 
the free energy, magnetization and internal energy of the Ising model
will be presented elsewhere \cite{MBBD08b}. 
We used several important
enhancements of the original Baxter approach \cite{BaxVar}, in particular an
improved iteration scheme \cite{Nish}, known as the corner transfer
matrix renormalization group (CTMRG). 
The calculations were performed  
for a grid of values of the magnetic field and temperature in the
range $10^{-7}<H<10^{-2}$ and $0.9 \beta_c<\beta<1.1\beta_c$, 
containing about $10,000$ distinct point (excluding a small region
around the critical point). 
The results for the scaling function $\Phi(\eta)$ are shown in
Fig.~\ref{figura1}. All calculated data points collapse on a smooth
curve, shown by the solid line (additional details presented on the
picture are explained below). Our numerical results for $\Phi(\eta)$ 
remarkably confirm 
the field theory calculations of \cite{FZ03}, to within all six significant
digits presented therein\footnote{%
We thank Alexander Zamolodchikov for providing us with additional
unpublished numerical data for $\Phi(\eta)$, which are again in perfect
agreement with our results.}.

The precision of our numerical calculations was tested against all
available exact 
results for the Ising model. In particular, the agreement between
calculated and exact values for the zero-field free energy \cite{O44},
magnetization \cite{O49} and magnetic susceptibility \cite{ONGP} in our
working range of temperatures varied  
between 14 and 28 decimal places (depending on the distance to the
critical point). In addition to these checks we also confirmed and extended 
many previously existing numerical results for the Ising model.  
Some details of our results are described below.

For the following discussion it is convenient to rewrite
\eqref{Phi-def} in an alternative form, introducing two more scaling
functions,
\be
{\mathcal F}(m,h)=\frac{m^2}{8\pi} \, \log m^2+
\left\{\begin{array}{ll}
m^2\,G_{high}(\xi), &\quad m<0\\
m^2\,G_{low}(\xi), &\quad m>0
\end{array}
\right.\ ,\qquad \xi={h}/{|m|^{15/8}}\ .\label{G-def}
\ee 
where $G_{high}(0)=G_{low}(0)=0$, corresponding to   
\be 
{\mathcal F}(m,0)=\frac{m^2}{8\pi} \, \log m^2\ .\label{Ft0}
\ee
These scaling functions are thoroughly discussed in \cite{FZ03}. 
The function $G_{high}(\xi)$ can be
expanded in a series in even powers of $\xi$
\be
G_{high}(\xi)=G_2 \xi^2 + G_4 \xi^4 +G_6 \xi^6 +\ldots \label{Ghigh-ser}
\ee
convergent in some domain around the origin of the $\xi$-plane.
The function $G_{low}(\xi)$ admits an asymptotic expansion 
\be
G_{low}(\xi)=\tilde{G}_1 \xi + \tilde{G}_2 \xi^2 +\tilde{G}_3 \xi^3
+\ldots
\label{Glow-ser} 
\ee
for small positive $\xi$. These new functions are simply related to
$\Phi(\eta)$. Note, 
in particular, that the coefficients $G_n$ and $\tilde{G}_n$
control the behavior of the function $\Phi(\eta)$ for large values of
$\eta$ on the real line,
\begin{eqnarray}
\Phi_{low}(\eta) &=& \tilde{G}_1 \eta^{\frac{1}{8}}
+ \tilde{G}_2 \eta^{-\frac{7}{4}} +\tilde{G}_3 \eta^{-\frac{29}{8}}
+\ldots
\,
\qquad \qquad \qquad {\rm for\ real}
\quad \eta \to +\infty\,,\label{Phi-low}\\[.3cm]
\Phi_{high}(\eta) &=& 
G_2 (-\eta)^{-\frac{7}{4}} + G_4 (-\eta)^{-\frac{22}{4}} +G_6
(-\eta)^{-\frac{37}{4}}
 +\ldots 
\quad {\rm for\ real}
\quad \eta \to -\infty\,.\label{Phi-high}
\end{eqnarray}
Finally, for small values of $\eta$, 
 \be
\Phi(\eta) = - {\frac{\eta^2}{8\pi}}\,\log\eta^2 + \sum_{k=0}^\infty
\Phi_k \eta^k \label{Phi-ser}
\ee
where the series converges in a finite domain around the origin 
of the complex $\eta$-plane.

Some of the above expansion coefficients are known exactly. 
The coefficient $\tilde G_1$ is known explicitly \cite{McCoyWu}
\be
\tilde G_1=-2^{1/12}\,e^{-1/8}\,{\mathcal A}^{3/2}=-1.357838341706595\ldots\ ,
\ee 
where ${\mathcal  A}=1.282427\ldots$ is the Glaisher constant.
The coefficients $G_2$ and $\tilde G_2$ have integral
expressions \cite{BMW,TM73} involving solutions of the Painlev\'e III
equation. They were numerically evaluated to very high precision
(50 digits) in \cite{ONGP}, 
\be
G_2=-1.845228078232838\ldots,\qquad \tilde G_2=-0.0489532897203\ldots\ .
\label{G2val}
\ee
The coefficient $\Phi_0$ was calculated in \cite{Fateev94},
\begin{equation}
\Phi_0=-\frac{(2\pi)^{\frac{1}{15}}\gamma(\frac{1}{3})
  \gamma(\frac{1}{5}) 
\gamma(\frac{7}{15})}
{\left[\gamma(\frac{1}{4}) \gamma^2(\frac{3}{16})\right]^\frac{8}{15}}=
-1.19773338379799\ldots ,\qquad \gamma(x)=\frac{\Gamma(x)}{\Gamma(1-x)}
\ee
where $\Gamma(x)$ is the standard $\Gamma$-function. The coefficient
  $\Phi_1$ has an explicit integral representation, obtained in
  \cite{FLZZ98}. We have evaluated the required integral explicitly,  
\be
\Phi_1=-\frac{  32 \cdot 2^\frac{3}{4}}{225\>(2\pi)^\frac{7}{15}}
\frac{\gamma(\frac{1}{3}) \gamma(\frac{1}{8}) \prod\limits_{k=3}^7
\gamma(\frac{k}{15})}
{\left[\gamma(\frac{1}{4})
  \gamma^2(\frac{3}{16})\right]^\frac{19}{15}}
=-0.3188101248906\ldots \ .\label{Phi1}
\end{equation}

Contrary to the field theory case, the lattice free energy,
\be
F(\tau,H)=F_{sing}(\tau,H)+F_{reg}(\tau,H)+F_{sub}(\tau,H), \qquad \tau,H\to0,
\label{Ffull}
\ee
never coincides with its leading universal part \eqref{fsing}.
It also contains regular terms $F_{reg}(\tau,H)$, analytic in $\tau$ and
$H$, as well as subleading singular terms $F_{sub}(\tau,H)$, 
which are non-analytic, but less singular than the first term in
\eqref{Ffull}.  
Therefore, to extract the universal scaling function 
from the lattice calculations one should be able to isolate and
subtract these extra terms. Moreover, one needs to
know the explicit form of the non-linear scaling variables
\eqref{mh-def}. In principle, all this information can be determined
entirely from numerical calculations (provided one assumes the values of
exponents of the subleading terms, predicted by the analysis
\cite{CPA02,ONGP} of
the CFT irrelevant operators, contributing to the free energy \eqref{Ffull}).
The accuracy of this ``fully numerical'' approach, however,
deteriorates rapidly for the higher order terms. Much more accurate
results can be obtained if the numerical work is combined   
with known exact results.

Write the non-linear variables  \eqref{mh-def} in the form,
\begin{eqnarray}
m(\tau,H)&=&-\sqrt{2}\, \tau \ a(\tau)+H^2\,b(\tau)+O(H^4)\ ,\nonumber\\[.3cm]
h(\tau,H)&=&C_h H\,\Big[c(\tau)+H^2\,d(\tau)+O(H^4)\Big]\ ,\label{mh-def1}
\end{eqnarray}
where $a(0)=c(0)=1$, $h(\tau,H)=-h(\tau,-H)$.
Similarly, write the
regular part as, 
\be
F_{reg}(\tau,H)=A(\tau)+H^2\,B(\tau)+O(H^4)\ .\label{Freg}
\ee
As shown in \cite{ONGP}, 
the most singular subleading term, contributing to \eqref{Ffull} is of
the order of $O(\tau^{9/4}\,H^2)$ (see Eq.\eqref{Fsub} below). The
fact that subleading terms arise only in such a high order ($\sim
m^6$) is a remarkable property of the square lattice nearest-neighbour
Ising model, which greatly simplifies the calculation of the universal
scaling function from the numerical data. 
  
For $H=0$ the expression \eqref{Ffull} should
reduce to Onsager's exact result \cite{O44}
\be
F(\tau,0)=\log \sqrt{2} \cosh (2\beta)+\int_0^{\pi}\frac{d\theta}{2\pi}\,\log\Big[
1+\Big(1-\frac{\cos^2\theta}{1+\tau^2}\Big)^{1/2}\Big]\ .
\ee
Therefore one should be able to rewrite the last formula in the form
\eqref{Ft0} plus regular terms. This is achieved by choosing \cite{Nick99}
\be
a(\tau)=\left[\int_0^1 d x\,
  F(\hf,\hf,1;-x\tau^2)/(1+x\tau^2)^{1/2}\right]^{1/2}=
{ 1-\frac{3\tau^2}{16}+\frac{137\tau^4}{1536}+
O(\tau^6)}\ ,\label{atau}
\ee
where $F(a,b,c,z)$ denotes the Gauss hypergeometric function. The
corresponding contribution to \eqref{Freg} then reads
\be
A(\tau)=-\frac{\ds 2{\mathcal G}}{\ds \pi}-\frac{\ds \log 2}{\ds 2}+
\frac{\ds \tau}{\ds 2}-\frac{\ds \tau^2(1+5\log 2)}{\ds 4\pi}-
\frac{\ds \tau^3}{\ds 12}+\frac{\ds 5\tau^4(1+6\log{2})}{\ds 64\pi}
+O(\tau^5)\ ,
\ee
where ${\mathcal G}=0.915965594\ldots$ is the Catalan constant.

Next, using the exact expression for the zero field magnetization
 \cite{O49,Y52}
\be
M(\tau,0)=(1-k(\tau)^2)^{1/8},\qquad \tau<0
\ee
with $k(\tau)$ defined in \eqref{k-def}, one finds from
\eqref{Z-def}, \eqref{G-def}, \eqref{Glow-ser} and \eqref{Ffull}
\be
c(\tau)=\frac{M(\tau)}{(-4\tau\,a(\tau))^{1/8}}
=1+\frac{\tau}{4}+
\frac{15\tau^2}{128}-\frac{9\tau^3}{512}-\frac{4333\tau^4}{98304}+O(\tau^5)
\ ,\label{ctau}
\ee 
and also 
\be
C_h=-2^{3/16}/\tilde{G}_1=0.838677624411\ldots\ .
\ee 

Finally, consider the zero-field susceptibility. 
 No simple closed form expression for the zero-field susceptibility
$\chi(\tau)$ is known. However, the authors of \cite{ONGP} obtained
remarkable asymptotic expansions of $\chi(\tau)$ for small
$\tau$ to within $O(\tau^{14})$ terms with high-precision numerical 
coefficients. We have compared their series
with our numerical results for the susceptibility in the range of
temperatures $|\tau|=0.05\mbox{--}0.14$ and found that they agree to each
other in 14 to 18 significant digits (depending on the value of
$\tau$). Their result can be written as (retaining the terms up to
 $O\big(|\tau|^{9/4}\big)$, inclusive)
\begin{eqnarray}
\chi(\tau)_{\rm ONGP}&=&-2^{-\frac{7}{8}}\, C_h^2 \, G''(0)\
|\tau|^{-\frac{7}{4}} 
\Big(1+\frac{\tau}{2}+\frac{5\tau^2}{8}+\frac{3\tau^3}{16}-
\frac{23\tau^4}{384}+O(\tau^5)\nonumber
\Big)\\[.3cm]
&&+e(\tau)+f(\tau)\log |\tau|+O(\tau^3 \log|\tau|)\label{sus2} 
\end{eqnarray}
where $e(\tau)$ and $f(\tau)$ are explicitly known second order
polynomials in $\tau$. The coefficient in front of the first term is
written in our notations\footnote{%
The correspondence with the notations  of ref.\cite{ONGP} is as
follows. Their $\beta \chi(\tau)$ is denoted here as 
$\chi(\tau)_{\rm ONGP}$. 
Their coefficients $C_\pm$ are connected to 
our constants by 
\be
C_+=-2^{\frac{1}{8}}\,\big(2\beta_c
\sqrt{2}\big)^{-\frac{7}{4}}\,C_h^2\, G_2\ ,\qquad
C_-=-2^{\frac{1}{8}}\,\big(2\beta_c
\sqrt{2}\big)^{-\frac{7}{4}}\,C_h^2\, \tilde{G}_2\ ,\nonumber
\ee
where $\beta_c$ is given in \eqref{beta-c}.}. 
The symbol $G''(0)$ there stands for the second derivative of the
scaling function  
$G(\xi)$ with respect to its argument. Namely  
$G''(0)=G_{high}''(0)=2G_2$ for $\tau>0$ and
 $G''(0)=G_{low}''(0)=2\tilde{G}_2$ for $\tau<0$. The corresponding
values are given in \eqref{G2val} above. 
Next, calculating the
second field derivative of \eqref{Ffull} at $H=0$, one obtains
\begin{eqnarray}
\chi(\tau)&=&-2^{-\frac{7}{8}}\, C_h^2 \, G''(0)\ |\tau|^{-\frac{7}{4}}
\, a(\tau)^{-\frac{7}{4}}\,c(\tau)^2 
-\frac{\partial^2 F_{sub}(\tau,H)}{\partial H^2}\Big\vert_{H=0}
\nonumber \\[.3cm]
&&-2B(\tau)+\frac{\tau\,a(\tau)\,b(\tau)}{\sqrt{2}\,\pi}
\Big(1+\log\big(2\tau^2 a(\tau)\big)\Big)\ . \label{sus} 
\end{eqnarray}
Equating this expression 
to \eqref{sus2} and  using \eqref{atau}, \eqref{ctau} and explicit
forms of the polynomials $e(\tau)$ and $f(\tau)$ from \cite{ONGP}, one
obtains 
\be
B(\tau)=0.0520666225469+0.0769120341893\, \tau+0.0360200462309\,
\tau^2+O(\tau^3)\label{Btau}\ ,
\ee 
and 
\be
b(\tau)=\mu_h\,\Big(1+\frac{\tau}{2}+
 O(\tau^2)\Big),\qquad \mu_h=0.071868670814\ .
\ee
Noting that
\be
a(\tau)^{-\frac{7}{4}}\,c(\tau)^2 =
1+\frac{\tau}{2}+\frac{5\tau^2}{8}+\frac{3\tau^3}{16}-
\frac{11\tau^4}{192}+O(\tau^5)
\ee
one immediately obtains the main contribution to the subleading term  
\be
\Big(2^{-\frac{7}{8}}\,C_h^2\,
  G''(0)\Big)^{-1}\,
\frac{\partial^2 F_{sub}(\tau,H)}{\partial H^2}\Big\vert_{H=0}
=-\frac{1}{384}\,\,|\tau|^{\frac{9}{4}} +\ldots \ .\label{Fsub}
\ee
The results of \cite{ONGP} provided the first convincing
demonstration of
the violation to simple one-parametric scaling in the square
lattice Ising model. Note that despite the $O(|\tau|^{9/4})$ 
term gives a very small contribution to the susceptibility we were
able to confidently quantify it from our numerical results. Namely, we
estimated the 
coefficient of the $\tau^4$ term in the series shown in parenthesis in the
first line of \eqref{sus2}. Our estimate is $23.004/384$ which is
within $0.02\%$ of its exact value $23/384$.

The coefficient $d(\tau)$ in \eqref{mh-def1} 
was estimated from our numerical data for the internal energy,
\be
d(\tau)=e_h+O(\tau),\qquad e_h=-0.007(1),
\ee
which is in agreement with the result $e_h=-0.00727(15)$ from \cite{VicC}. 

The above expressions were used to analyze our extensive numerical data and
extract the necessary information to obtain the universal scaling function. The
results are summarized in four tables. For convenience of
comparison we quoted the corresponding results from \cite{FZ03},
including those obtained through the TFFSA, the high/low-temperature 
and extended dispersion relations (DR). Earlier 
exact and numerical results for the
same quantities are also quoted (whenever available). 
Figure~1 
shows about 10,000 data points for the scaling function $\Phi(\eta)$. 
As expected, the points collapse on a smooth curve (their spread is
much smaller than  the resolution of the picture). Figure~1 also shows plots of
the asymptotic expansions \eqref{Phi-low}, \eqref{Phi-high} and
\eqref{Phi-ser} with the coefficients given in Table~1, Table~2 and Table~3.
These expansions are seen to ``stitch'' together very well and give a 
reasonably good analytic
approximation to $\Phi(\eta)$ in the whole real line of $\eta$.
Our Fig.~\ref{figura1} essentially coincides with Fig.~10 of
\cite{FZ03}. The numerical values of $\Phi(\eta)$ at small integer
values of $\eta$ are given in Table~4.

The calculations were performed on the 24-processor Linux Cluster
System at the ANU Research School of Physics and Engineering and 
on the 1928-processor SGI Altix 3700 Bx2 Cluster at the ANU 
Supercomputer Facility. The level of parallelization varied between 15
and 50 processors.  The total amount of CPU time spent for this
work was about 9,000 hours (single processor equivalent).

To conclude, we have implemented Baxter's variational corner transfer
matrix approach to obtain the universal scaling function of the square
lattice Ising model in a magnetic field, as shown in Figure~1 and
Table~4.
The numerical data is seen to be in remarkable agreement with the field theory
results obtained by Fonseca and Zamolodchikov \cite{FZ03}.
We also report a remarkable agreement (11 to 14 digits) between 
our numerical values for $\tilde{G}_1$,  $G_2$ and 
$\tilde{G}_2$ and the classic exact results of Barouch, McCoy, Tracy and
Wu \cite{McCoyWu,BMW,TM73} (see Tables 1 and 2), and a similar
agreement between the values $\Phi_0$ and $\Phi_1$ and the exact predictions
\cite{Fateev94,FLZZ98} of Zamolodchikov's integrable $E_8$ field
theory \cite{Zam89a} (see Table~3).
The high precision of
the numerical results underline the full power and further potential of
the variational corner transfer matrix approach. 
In this case, the results
show beyond any doubt the validity of the connection between the scaling
limit of the Ising model in a magnetic field and the Ising Field Theory 
\eqref{IFT}.

The authors thank H.~Au-Yang, R.J.~Baxter, G.~Delfino, 
S.L.~Lukyanov, J.H.H.~Perk,
C.A.~Tracy and
A.B.~Zamolodchikov for useful discussions and remarks.
This work has been partially supported by the Australian Research Council.



\newcommand\oneletter[1]{#1}

\vspace{3cm}

\begin{table}[p] 
\centering
\begin{tabular}{l|l|l|ll}
& \multicolumn{1}{|c|}{CTM (This work)} 
& \multicolumn{1}{|c}{High-$T$ DR \cite{FZ03}}
& \multicolumn{2}{|c}{From References}\\
\hline
$G_2$ \rule{0cm}{.4cm}& $-1.8452280782328(2)$  & $-1.8452283$& $-1.845228078232838...$ &\cite{BMW,TM73,ONGP} \\
$G_4$ & $\phantom{+}8.333711750(5)$ & $\phantom{+}8.33410$& $\phantom{+}8.33370(1)$&\cite{VicC}\\
$G_6$ & $-95.16896(1)$ & $-95.1884$& $-95.1689(4) $&\cite{VicC}\\
$G_8$ & $\phantom{+}1457.62(3)$ & $\phantom{+}1458.21$& $\phantom{+}1457.55(11)$&\cite{VicC}\\
$G_{10}$ & $-25891(2)$ & $-25889$ & $-25884(13)$&\cite{VicC}\\
\end{tabular}

\caption{
Numerical values of the coefficients $G_{2n}$ in (\ref{Ghigh-ser}).
The second column contains the high-temperature dispersion relation  (DR)
results from \cite{FZ03}. The coefficient $G_2$ is known exactly
\cite{BMW,TM73,ONGP}.
} 
\end{table}


\begin{table}[p]
\centering
\vspace{1.cm}
\begin{tabular}{l|l|l|l}
& \multicolumn{1}{|c|}{CTM (This work)} 
& \multicolumn{1}{|c}{Low-$T$ DR \cite{FZ03}} 
& \multicolumn{1}{|c}{From References} \\
\hline
$\tilde G_1$\rule{0cm}{.4cm} & $-1.3578383417066(1)$  & $-1.35783835$ & $-1.357838341706595...$  \cite{McCoyWu}\\
$\tilde G_2$ & $-0.048953289720(1)$ & $-0.0489589$ & $-0.0489532897203... $  \cite{BMW,TM73,ONGP} \\
$\tilde G_3$ & $ \phantom{+} 0.038863932(3)$ & $\phantom{+} 0.0388954$& $\phantom{+}0.0387529$ \cite{MW78}; 
$\phantom{+}0.039(1)$ \cite{Zin96}\\
$\tilde G_4$ & $-0.068362119(2)$ & $-0.0685060$& $ -0.0685535$ \cite{MW78}; $-0.0685(2)$ \cite{Zin96} \\
$\tilde G_5$ &  $\phantom{+}0.18388370(1)$  & $\phantom{+}0.18453$ & \multicolumn{1}{|c}{---}\\
$\tilde G_6$ &  $-0.6591714(1)$ & $-0.66215$& \multicolumn{1}{|c}{---}\\
$\tilde G_7$ &  $\phantom{+}2.937665(3)$& $\phantom{+}2.952$&\multicolumn{1}{|c}{---}\\
$\tilde G_8$ &  $-15.61(1)$ & $-15.69$&\multicolumn{1}{|c}{---}\\
\end{tabular}

\caption{
Numerical values of the coefficients $\tilde G_{n}$ in (\ref{Glow-ser}).
The second column contains the low-temperature dispersion relation results
from \cite{FZ03}. The right column refers to exact values of
$\tilde{G}_1$ \cite{McCoyWu} and $\tilde{G}_2$ \cite{BMW,TM73,ONGP} and
other numerical results.
}
\vspace{1.cm}
\end{table}

\clearpage

\begin{table}[p]
\centering
\vspace{1.cm}
\begin{tabular}{l|l|l|l|ll}
& \multicolumn{1}{|c|}{CTM (This work)} 
& \multicolumn{1}{|c|}{TFFSA \cite{FZ03}} 
& \multicolumn{1}{|c|}{Ext. DR \cite{FZ03} } 
& \multicolumn{1}{|c}{From References} \\
\hline
 $\Phi_0$ \rule{0cm}{.4cm}& $-1.197733383797993(1)$  
& $-1.1977331$ & $-1.1977320$ & $-1.19773338379799339...$  \cite{Fateev94}\\
$\Phi_1$ & $-0.318810124891(1)  $ & $-0.3188103$ & $-0.3188192$ &
$-0.3188101248906...$  \cite{FLZZ98}\\ 
$\Phi_2$& $\phantom{+}0.110886196683(2)$ & $\phantom{+}0.1108867$ &
$\phantom{+}0.1108915$ & \multicolumn{1}{|c}{---}\\ 
$\Phi_3$& $\phantom{+}0.01642689465(2)$ & $\phantom{+}0.0164266$ & $\phantom{+}0.0164252$ & \multicolumn{1}{|c}{---}\\
$\Phi_4$& $-2.639978(1)\times10^{-4}$ & $-2.64\times10^{-4}$ &$-2.64\times10^{-4}$ & \multicolumn{1}{|c}{---}\\
$\Phi_5$& $-5.140526(1)\times10^{-4}$ & $-5.14\times10^{-4}$ & $-5.14\times10^{-4}$ & \multicolumn{1}{|c}{---}\\
$\Phi_6$& $\phantom{+}2.08865(1)\times10^{-4}$ & $\phantom{+}2.07\times10^{-4}$ & $\phantom{+}2.09\times10^{-4}$ & \multicolumn{1}{|c}{---}\\
$\Phi_7$& $-4.4819(1)\times 10^{-5}$ &  $-4.52\times10^{-5}$ & $-4.48\times10^{-5}$& \multicolumn{1}{|c}{---}\\
$\Phi_8$& \multicolumn{1}{|c|}{---} & \multicolumn{1}{|c|}{---} & $\phantom{+}3.16\times10^{-7}$&\multicolumn{1}{|c}{---}\\
$\Phi_9$& \multicolumn{1}{|c|}{---} &  \multicolumn{1}{|c|}{---} & $\phantom{+}4.31\times10^{-6}$&\multicolumn{1}{|c}{---}\\
$\Phi_{10}$& \multicolumn{1}{|c|}{---} & \multicolumn{1}{|c|}{---} &$-1.99\times10^{-6}$&\multicolumn{1}{|c}{---}\\
\end{tabular}
\caption{
Numerical values of the coefficients $\Phi_{n}$ in (\ref{Phi-ser}).
The second and third columns contain results from \cite{FZ03}, obtained
through the TFFSA and the extended dispersion relations (DR), respectively. 
The forth column refers to exact results; the numerical value 
of $\Phi_1$ therein is taken from Eq.\eqref{Phi1}. 
} 
\vspace{1.cm}
\end{table}

\begin{table}[p]
\centering
\begin{tabular}{l|c|c|c|c}
$\Phi(\eta)$ &  \multicolumn{1}{|c|}{CTM (This work)} & 
\multicolumn{1}{|c|}{TFFSA \cite{FZ03}} &
\multicolumn{1}{|c|}{High/Low-$T$ DR \cite{FZ03}}  &
\multicolumn{1}{|c}{Ext. DR \cite{FZ03}}  \\
\hline
$\Phi(-5)$& $-0.10920919$ &$-0.1092101$& $-0.1092092$ & $-0.1088626$\\
$\Phi(-4)$& $-0.15926438$ &$-0.1592682$& $-0.1592643$ & $-0.1589421$\\
$\Phi(-3)$& $-0.25298908$ &$-0.2529928$& $-0.2529887$ & $-0.2527417$\\
$\Phi(-2)$& $-0.44132564$ &$-0.4413450$& $-0.4413249$ & $-0.4412136$\\
$\Phi(-1)$& $-0.78396650$ &$-0.7839665$& $-0.7839668$ & $-0.7839576$\\
$\Phi(0)$ & $-1.19773338$ &$-1.1977330$&  --- & $-1.1977320$\\
$\Phi(1)$ & $-1.38984135$ &$-1.3898410$&   $-1.3898417$ & $-1.3898063$\\
$\Phi(2)$ & $-1.49305602$ &$-1.4930558$&   $-1.4930566$ & $-1.4929849$\\
$\Phi(3)$ & $-1.56427320$ &$-1.5642732$&   $-1.5642736$ & $-1.5641727$\\
$\Phi(4)$ & $-1.61885066$ &$-1.6188506$&   $-1.6188510$ & $-1.6187275$\\
$\Phi(5)$ & $-1.66324828$ &$-1.6632483$&   $-1.6632485$ & $-1.6631076$\\
\end{tabular}

\caption{
Numerical values of $\Phi(\eta)$ at small integer values of $\eta$.
The second, third and fourth columns contain results \cite{FZ03}
from the TFFSA, the high/low-temperature dispersion and the extended
dispersion relations (DR).
} 
\end{table}

\end{document}